\documentclass[12pt]{article}
\usepackage{times}
\usepackage{geometry}
\usepackage[utf8]{inputenc}
\usepackage{enumitem,amssymb}
\usepackage{ragged2e}
\usepackage{xcolor,wasysym,natbib,hyperref,graphicx,wrapfig}
\newlist{thematic}{itemize}{8}
\setlist[thematic]{label=$\square$}
\usepackage{pifont}

\begin{document}
\begin{flushleft}
\huge
Astro2020 Science White Paper \linebreak

High-Dimensional Dust Mapping \linebreak
\normalsize

\noindent \textbf{Thematic Areas:} \hspace*{60pt} $\square$ Planetary Systems \hspace*{10pt} $\XBox$ Star and Planet Formation \hspace*{20pt}\linebreak
$\square$ Formation and Evolution of Compact Objects \hspace*{31pt} $\square$ Cosmology and Fundamental Physics \linebreak
  $\square$  Stars and Stellar Evolution \hspace*{1pt} $\XBox$ Resolved Stellar Populations and their Environments \hspace*{40pt} \linebreak
  $\XBox$    Galaxy Evolution   \hspace*{45pt} $\square$             Multi-Messenger Astronomy and Astrophysics \hspace*{65pt} \linebreak
 

\textbf{Principal Author:}

Name: Gail Zasowski
 \linebreak						
Institution: University of Utah
 \linebreak
Email: gail.zasowski@gmail.com
 \linebreak 
Phone: $+$1-801-581-6901
 \linebreak
 
\textbf{Co-authors:} (names and institutions)
  \linebreak
(alphabetical) \\
Douglas~P.~Finkbeiner~(Harvard),
Gregory~M.~Green~(Stanford/KIPAC),
Juna~A.~Kollmeier~(OCIS),
David~M.~Nataf~(JHU),
J.~E.~G.~Peek~(STScI),
Edward~Schlafly~(LBNL),
Victor~Silva~Aguirre~(Aarhus),
Joshua~S.~Speagle~(Harvard),
Kirill~Tchernyshyov~(JHU),
Juan~D.~Trujillo~(UW),
Catharine Zucker~(Harvard)
\linebreak

\textbf{Abstract  (optional):}
Galactic interstellar dust has a profound impact not only on our observations of objects throughout the Universe, but also on the morphology, star formation, and chemical evolution of the Galaxy.  The advent of massive imaging and spectroscopic surveys (particularly in the IR) places us on the threshold of being able to map the properties and dynamics of dust and the interstellar medium (ISM) in three dimensions throughout the Milky Way disk and bulge.  These developments will enable a fundamentally new understanding of dust properties, including how grains respond to their local environment and how those environments affect dust attenuation of background objects of interest.  Distance-resolved maps of dust motion also hold great promise for tracing the flow of interstellar material throughout the Galaxy on a variety of scales, from bar-streaming motions to the collapse and dissolution of individual molecular clouds.  These advances require optical and IR imaging of stars throughout the Galactic midplane, stretching many kiloparsecs from the Sun, matched with very dense spectroscopic coverage to probe the ISM's fine-grained structure.

\end{flushleft}

\pagebreak

\setlength{\parindent}{16pt}
\section{Motivation}
\label{sec:intro}

The study of interstellar dust touches nearly all areas of astrophysics, from correcting reddened observations of background objects to following the path of heavy elements in and out of the gas-phase interstellar medium (ISM).  The ISM is the transit system by which all metals are carried from their formation sites throughout the galaxy and incorporated into later generations of stars and planets. Understanding how that transit system functions, the role that dust (the most metal-rich component of a galaxy) plays in it, and how dust varies spatially in the ISM has implications for galactic chemical evolution, star formation processes and regulation, and even planetary properties \citep[e.g.,][]{Unterborn_2017_exoplanetgeology}.

The majority of Milky Way (MW) dust studies fall into one of two categories: those seeking to understand detailed grain properties and chemistry, and those seeking to understand the on-sky distribution of the grains (or at least of their obscuring effect).  The first category of work frequently focuses on carefully-selected spatial regions at high resolution: e.g., multi-element grain depletion studies in the local ISM \citep[e.g.,][]{Jenkins_2019_OGeKrDepletions}, or variations in extinction law or opacity versus integrated $A_V$ in dense clouds \citep[e.g.,][]{Ascenso_2013_PipeExtLaw,Juvela_2015_ColdCoreOpacity}. The second category includes dust column density maps on scales spanning individual molecular clouds \citep[e.g.,][]{RomanZuniga_2009_B59extmap} up to the entire sky \citep[e.g.,][]{Green_2015_PS1dustmap}.  Both categories provide insights into dust populations that can be applied to more distant extragalactic systems.

Historically, most dust mapping has been in two dimensions, which has limited our ability to understand the {\it line-of-sight} variations in the dust properties that are known to vary tangentially {\it across} the sky.  With the advent of enormous photometric surveys and improvements in stellar distance measurements (e.g., from {\it Gaia}), highly sophisticated 3D maps of dust column density can be constructed.  Such 3D maps will enable the next leap in our understanding of dust in the Universe: {\bf the 3D-resolved distribution of interstellar dust properties, including attenuation behavior and projected velocity}.  These distributions are crucial for answering some of the key open questions in galactic astronomy: What physical processes impact grain size distribution, and in turn, how does the grain size distribution impact the ISM's ability to form stars, support chemical reactions, and modify incident starlight (\S\ref{sec:properties})?  What is the life-cycle of a typical molecular cloud, and what factors determines when it does and does not form stars (\S\ref{sec:small_flows})?  How is the ISM {\it flowing} throughout its galaxy, and how do those flows affect the chemical distributions of later stellar generations (\S\ref{sec:large_flows})?  
In this white paper, we provide motivation and examples of the impact that 3D-resolved maps of dust properties and dynamics will have on these critical questions.  The ability to construct these maps relies not only on the vast troves of stellar photometric and spectroscopic data being assembled now but also on future capabilities beyond the 2020 horizon (\S\ref{sec:future}).

\section{Dust Grain Properties}
\label{sec:properties}

\underline{\bf Extinction Curve and Grain Environment}:
The properties of dust grains vary with their environment, changing their interaction with radiation.  Surveys of ultraviolet dust extinction curves in the MW \citep[e.g.,][]{Fitzpatrick_2007}, in the Magellanic Clouds \citep[e.g.,][]{Gordon_2003}, and in M31 \citep{Clayton_2015} show substantial differences.  These changes in the extinction curve shape are tied to changes in the distribution of dust grain sizes and the properties of those grains, which in turn are determined by the environment of the grains and their history in the ISM.  However, the connection between the observed extinction curves and the underlying astrophysics is poorly understood.  New insights in this area are critical for understanding dust in other galaxies, especially at high redshift where measurement of the extinction curve on small spatial scales is impossible.

Properties of dust are often assessed through the parameter $R_V = A_V/E(B-V)$, which measures the steepness of the extinction curve in the optical, and which appears to encode much of the variation between different extinction curves \citep[e.g.,][]{Cardelli_1989}.  
The extinction curve and the parameter $R(V)$ can be measured using spectroscopy and photometry; spectroscopic temperature and metallicity measurements predict stars' intrinsic colors, which in comparison with observed colors give the extinction curve and $R(V)$.
Measurements of $R_V$ suggest tantalizing physical relationships between the dust and its environs.  However, existing studies are limited because their extinction curve measurements sample different environments very sparsely, making the connection between observed $A_\lambda$ variations and the physical mechanisms difficult to unravel.  For example, the extinction curve in the Small Magellanic Cloud features a less prominent 2175{\AA} bump than in the Milky Way, but it is not clear whether this is due to the lower metallicity of the Small Magellanic Cloud or to its greater star-formation activity \citep{Gordon_2003}.  Current imaging surveys (such as PanSTARRS and 2MASS) combined with current spectroscopic surveys (such as APOGEE) allow the variation in $R_V$ in the nearby Milky Way ($D < 4$ kpc) to be clearly resolved for stars with $A_V > 1$ \citep[e.g.,][]{Schlafly_2016_OIRextinctionlaw}, but at too coarse a resolution to allow these ambiguities to be resolved.

Modern spectroscopic surveys of millions of stars provide a route forward.  Until the 2010s, the stars with precise $R_V$ measurements numbered in the hundreds.  Recent work has expanded that number a hundred-fold, revealing that $R_V$ (presumably tracking the grain size distribution) is strongly correlated with the shape of the far-IR dust spectral energy distribution, and that most of the variation in $R_V$ observed for typical stars in the Milky Way is not associated with the present density of the local interstellar medium \citep{Schlafly_2016_OIRextinctionlaw, Schlafly_2017}. These observations suggest that dust properties may reflect the grains' earlier history as well as their present situation, and that simple ISM diagnostics (like density) may be inadequate proxies for predicting dust extinction. 
In concert with {\it Gaia} parallaxes and improved estimates of stars' absolute magnitudes from their spectra \citep[e.g.,][]{Hogg_2018, Leung_2019}, the properties of dust can be probed in three dimensions.  Access to the third dimension is critical, since it allows dust properties to be correlated with the local ISM {\it volume} density and radiation field strength.

Measuring the dust properties via the extinction curve at high resolution in 3D will link the extinction curve to the physical properties that determine it.  In particular, we will be able to address the following questions:
\vspace{-8pt}
\begin{itemize} \itemsep -3pt
    \item Do dust grains in dense regions of the ISM coagulate, reducing the number of small dust grains and leading to flatter extinction curves?
    \item Do shocked regions of the ISM shatter dust grains and steepen the extinction curve---for instance, around supernovae?
    \item How do other properties of the interstellar environment, like the metallicity of the gas and the spectrum of the radiation field, affect the properties of the dust grains?
\end{itemize}
Resolving these issues requires dense sampling of stars in the Galactic disk with optical photometry and optical or IR spectroscopy to provide stellar parameters and intrinsic colors (\S\ref{sec:future}).

\noindent\underline{\bf Infrared Extinction Curve}:
All of the questions outlined above apply to the extinction curve at IR wavelengths as well.
It was once (and sometimes still is) assumed that IR extinction universally follows a form $A_{\lambda} \propto \lambda ^ {-1.61}$; however, it has now been measured to vary dramatically, even on small angular scales, with the steepness reaching $A_{\lambda} \propto \lambda ^ {-2.47}$ \citep{AlonsoGarcia_2017_VVVextlaw}.
IR observations will be the predominant observing mode of a large fraction of the next large facilties and missions (e.g., {\it JWST}, SDSS-V, MOONS, {\it WFIRST}, TMT, GMT), including those targeting Type~Ia~SNe and other tracers of precision cosmology that require extremely precise extinction maps \citep{Huterer13}.  Though extinction corrections towards cosmological sources rely on integrated extinction maps, not distance-resolved ones, understanding the 3D dust environment along each line of sight remains critical to predicting the extinction curve. 
Densely packed spectroscopy of stars with multi-band IR photometry (\S\ref{sec:future}) is essential to characterize the IR extinction curve and its variation well enough to fully exploit the tremendous investments in next-generation IR observations. 


\noindent\underline{\bf Dust and the DIBs}:
The relationship between dust absorption and the diffuse interstellar bands \citep[DIBs;][]{Cox_2011_DIBssummary} offers another potential window into the 3D dust environment. Firstly, observations of sightlines with detections of multiple DIBs and constraints on other ISM diagnostics have demonstrated that as a family, the DIBs are highly responsive to local ISM conditions \citep[e.g.,][]{Kos_2013_dibsphysconditions}.  This sensitivity means that DIB measurements are an effective tool for probing the local properties of ISM at different locations in the Galaxy.  Secondly, certain DIBs that are particularly well-correlated with dust extinction are powerful, independent probes of foreground-only attenuation, even in dusty parts of the Galaxy without high-quality 3D dust maps \citep[e.g.,][]{Munari_2000_gaiaDIB,Zasowski_2015_dibs,DeSilva_2015_galah}. Thirdly, while radial velocities can't be measured directly for dust clouds, they {\it can} be measured for these spectral features, providing a unique means of tracing the foreground dust kinematics of any cloud between the Sun and a background source (see \S\ref{sec:small_flows}--\ref{sec:large_flows}). 
Integrating these observations together to extract information on the ISM properties and phase space requires understanding how the correlation between different DIB equivalent widths ($W_{\rm DIB}$) and $A_V$ depends on ISM environment.  
These measurements require wide-wavelength spectroscopy to probe $W_{\rm DIB}/A_V$ for different DIBs, especially $H$-band spectra to measure the well-correlated 1.53~$\mu$m DIB in high-extinction regions \citep{Zasowski_2015_dibs}, along with optical photometry and sufficient stellar information (from spectroscopy and/or wide-baseline photometry) to estimate $A_V$ (\S\ref{sec:future}).


Even today's surveys that dwarf earlier extinction law measurements still sparsely sample the Galactic disk, relative to the scale of the ISM's 3D fluctuations. Future, {\it denser} surveys will be able to obtain much higher resolution throughout the MW better matched to the ISM's behavior (\S\ref{sec:future}).


\section{Small Scale Flows and Star Formation}
\label{sec:small_flows}
One critical step on the path to star formation is the evolution of the star formation sites---the collapse and fragmentation of molecular clouds.
Why and how do certain patches of warm diffuse gas become patches of cold dense gas, able to fragment into super-dense star-forming cores?
The proliferation of analytical models and simulations \citep[e.g.][]{Clark12,C-N14, Heitsch13} predict a wide range of mechanisms and outcomes, including clouds that form quickly through colliding ISM flows, clouds that collapse slowly via a semi-stochastic accumulation of matter, and clouds that are stable over long timescales.

It is important to be able to distinguish between these scenarios because they correspond to different pictures of where star formation can happen, what parts of the cloud and star formation process can be detected with different tracers, and how efficient feedback processes need to be at disrupting molecular clouds in order to balance their formation rate.
Clouds formed from colliding flows can only form given a specific 
dynamical impetus, may start forming stars before they are clearly detectable in CO, 
and will be short-lived. 
Conversely, clouds that collapse semi-stochastically can form in any
turbulent environment, will form CO
before commencing star formation, and may live significantly longer.  

Distinguishing between the scenarios requires accurate measurements of either cloud lifetimes or the actual gas flows around still-forming molecular clouds.
Existing observations have focused most heavily on measuring cloud lifetimes from population statistics of tracers of different stages of cloud and star formation \citep[e.g.][]{VS18}.
However, so far this type of analysis has not been able to conclusively determine the predominant cloud formation process.



\begin{wrapfigure}[20]{r}{3.5in}
\begin{center}
\begin{minipage}{3.5in}
\begin{center}
\vskip -0.8cm
    \includegraphics[width=0.95\textwidth]{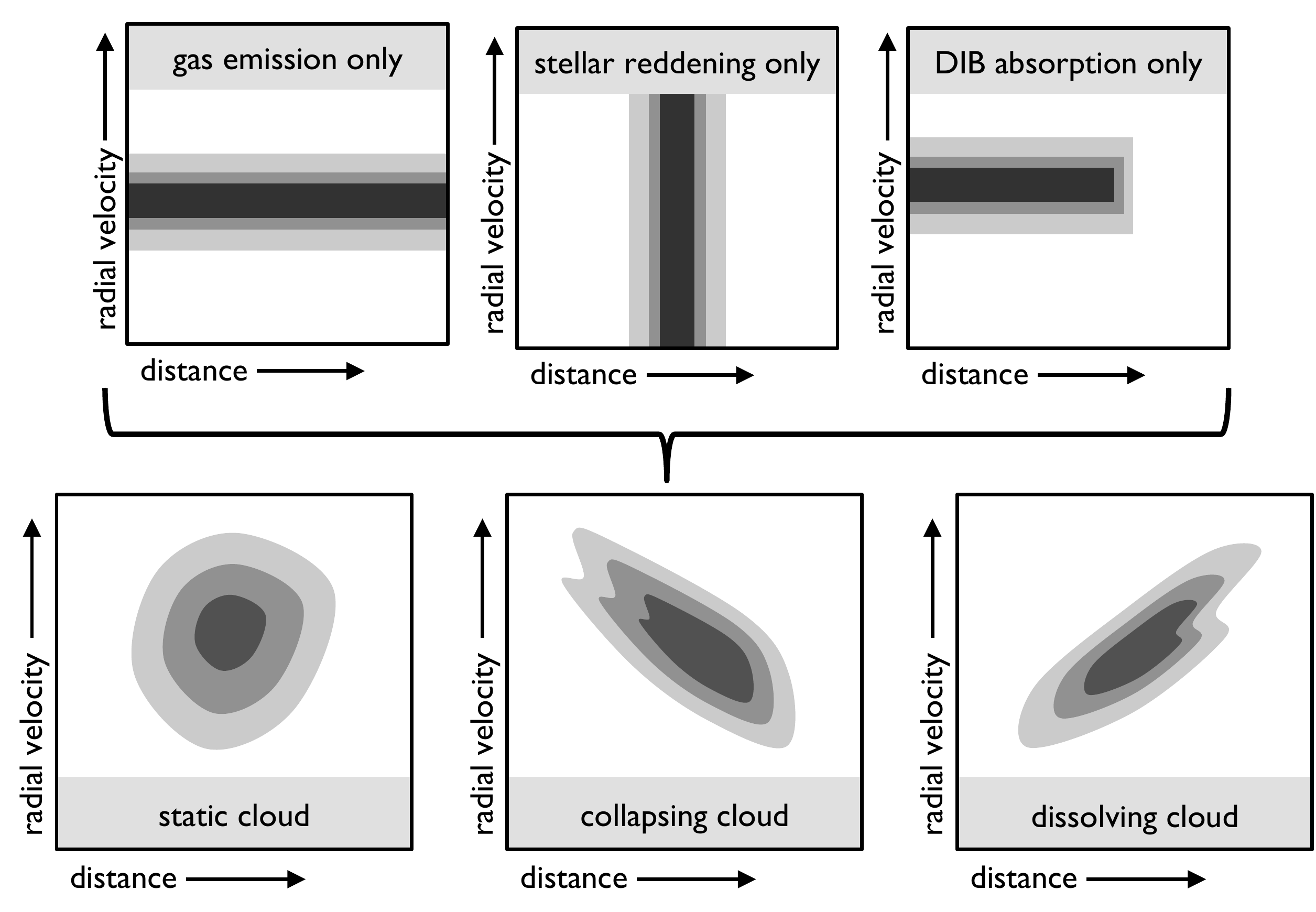}
    \caption{\textcolor{darkgray}{Cartoon demonstrating the power of combining ISM velocity and distance information. {\bf Top:} Gas emission, DIB absorption, and stellar reddening measurements constrain different dimensions of the ISM's spatial--kinematic structure. {\bf Bottom:} This spatial--kinematic structure reveals the cloud's dynamical status.}
    }
    \label{fig:kt_cartoon}\end{center}
\end{minipage}
\end{center}
\end{wrapfigure}
One powerful approach to measuring the flow of gas onto a still-forming molecular cloud is to directly map the line-of-sight velocity of the ISM as a function of three-dimensional position.  By employing different combinations of independent velocity- and distance-resolved ISM measurements (Figure~\ref{fig:kt_cartoon}), this ``kinetic tomography'' method has been applied to measure large-scale velocity fields across many kpc of the Galactic disk \citep[\S\ref{sec:large_flows}; e.g.,][]{Tchernyshyov_2017_kt1,Tchernyshyov_2018_kt2}. If adapted for focused observations of individual molecular clouds, with sufficient background stellar target density and precise (e.g., {\it Gaia}) distances, this method could be used to map the infalling or outflowing motions of a molecular cloud's diffuse gas envelope.  A large survey of multiple cloud envelopes in the solar neighborhood and in different Galactic environments (e.g., in and out of spiral arms) would elegantly trace the life-cycle of molecular clouds and the factors that drive their star formation (\S\ref{sec:future}). 

\section{Large Scale Flows}
\label{sec:large_flows}

The global velocity field of a galaxy's ISM reflects the galaxy's graviational potential and its recent interaction history.  
Because the ISM is dynamically cold, its strong response to perturbations make it a useful tracer both on its own and when compared with dynamically hotter tracers (e.g., older stars).
For example, \citet{Tchernyshyov_2018_kt2} showed that the MW's distance-resolved velocity field could be used to distinguish between different spiral arm formation scenarios that produced very similar morphological structures.  In M31, \citet{Quirk_2019_M31asymmetricdrift} used gas rotation observations to help constrain the dynamical heating of the disk stellar populations.  The damping of the ISM's response to transient perturbations means that comparisons between stellar and gas kinematics can also place constraints on the age and timescale of interactions \citep[e.g.,][]{Lipnicky_2018_satelliteHIperturb}.  These types of analyses are more common in external galaxies, where velocity and (2-D projected) position are more easily simultaneously obtained, and highlight the analytic power that could be brought to MW studies.



Another regime where distance-resolved ISM dynamics will be immensely valuable is in understanding accretion and feedback at the Galactic disk/halo interface.  Observational evidence, such as the presence of (relatively) enriched gas in the high halo 
\citep{2011ApJ...739..105S} and the need for ``leaky boxes'' to model stellar metallicity distributions \citep[e.g.,][]{Chiappini97}, suggests that interstellar flows must bring gas and dust into the Galaxy at some points and launch it outwards at others.  Both processes have largely been studied and modeled in external systems, with a focus on particularly energetic launching points and on flows in the circumgalactic medium \citep[e.g.,][]{Somerville_2015_galaxyformation}.  Kinetic tomography holds great promise for directly inferring ISM motions in the thick disk and inner halo, where there are sufficient stars to provide an estimate of the clouds' spatial distribution. Studying these feedback and accretion events in the MW will enable us to understand not only their effects in a 3D spatially resolved way, but also how less-energetic energy sources contribute to the global ISM flow patterns.

Obtaining ISM velocities using DIB features requires high-resolution spectroscopy, so that multiple cloud components along a given line of sight can be separated, and a robust determination of stellar parameters (or a well-characterized set of DIB-free spectra) so that the stellar contribution can be cleanly subtracted.  Like \S\ref{sec:properties} and \S\ref{sec:small_flows} above, this approach requires dense sampling of the Galactic plane in order to resolve distant flows.

\section{Future Requirements}
\label{sec:future}

The ISM features physically important variations in its dust properties and gas velocities over small spatial scales \citep[e.g.,][]{Nataf_2013_bulgeextlaw,Schlafly_2016_OIRextinctionlaw}. {\it To accurately measure the 3-space distribution of these quantities and how they relate to the structure of the ISM across the disk, dense sampling of a wide range of environments is needed, via infrared photometric and spectroscopic observations of millions of stars.} 

Thoroughly resolving the questions posed in \S\ref{sec:properties}--\ref{sec:large_flows} requires optical $+$ IR photometry and near-IR spectroscopy (at moderate-to-high resolution) for tens to hundreds of stars per (100 pc)$^3$ volume element throughout the Galaxy.  This target density is necessary since 100~pc is roughly the typical separation between molecular clouds and the size of supernovae-driven bubbles in the ISM.  This motivates observations of millions of stars in the disk within several kpc of the Sun.  Existing and imminent surveys of the Galactic plane will start to provide the needed observations over the next decade, but further investment in massive IR spectroscopy of stars in the Galactic disk is needed to fully realize the potential of this program.  The next frontier will be extending matched imaging$+$spectroscopic observations to stars deep in the inner Galactic disk and bulge, where upcoming surveys are too shallow to obtain the precision and target density needed and where {\it Gaia} cannot penetrate the obscuring dust.  
Dramatic advances in our knowledge of the 3D properties of the ISM are possible only with continued investment in spectroscopic, photometric, and astrometric measurements of dense samples of millions of stars throughout the Milky Way.

\pagebreak

\bibliographystyle{aa}
\bibliography{ms}

\end{document}